\documentclass[aps,prl,twocolumn,showpacs,letterpaper,superscriptaddress]{revtex4}
\usepackage{latexsym}
\usepackage{amsmath}
\usepackage{mathrsfs}

\newcommand{\lagr}{\mathscr{L}}
\newcommand{\hamil}{\mathscr{H}}

\begin{document}
\title{Condensed matter many-body theory in relativistic covariant form}

\author{Valerio \surname{Olevano}}
\affiliation{Institut N\'eel, CNRS \& UJF, Grenoble, France}
\affiliation{Istituto di Cristallografia, CNR, Bari, Italy}
\affiliation{European Theoretical Spectroscopy Facility (ETSF)}
\author{Massimo \surname{Ladisa}}
\affiliation{Institut N\'eel, CNRS \& UJF, Grenoble, France}
\affiliation{Istituto di Cristallografia, CNR, Bari, Italy}
\affiliation{European Theoretical Spectroscopy Facility (ETSF)}
\date{\today}
\pacs{03., 11.10.-z, 12.20.-m, 71.10.-w}

\begin{abstract}
We present a relativistic covariant form of many-body theory.
The many-body covariant Lagrangian is derived from QED by integrating out the internal non-quantized electromagnetic field.
The ordinary many-body Hamiltonian is recovered as an approximation to the exact covariant theory that contains many-body terms beyond the solely electrostatic interaction, e.g. the Lorentz force among electrons, spin-spin etc.
Spin and relativistic terms, e.g. spin-orbit, are also automatically accounted.
Moreover, the theory is compact, gauge-invariant and respects causality.
\end{abstract}

\maketitle

\paragraph{Introduction}
The physical world consists of interacting many-body systems.
The description of a many-body system is a problem even in classical mechanics and becomes a formidable problem in quantum mechanics.
The direct solution of the Schr\"odinger equation is already unfeasible for $N>10$ particles and the N-body wavefunction is impractical for $N \sim 10^{23}$, as in condensed matter.
One has to resort to other formalisms.
Many-body (MB) quantum field theory (QFT) \cite{FetterWalecka,Mahan,BaymKadanoff}, better known as many-body perturbation theory (MBPT), was historically the formalism to tackle the many-body problem.
As any other QFT, it relies on second quantization and the use of Green's functions.
MBQFT was pioneered by Baym, Kadanoff \cite{BaymKadanoff}, Hedin \cite{Hedin} and Keldysh, relying on previous work on quantum electrodynamics (QED) mainly by Feynman, Schwinger and Dyson 
\cite{DysonPR75_1,DysonPR75_2,SchwingerPNAS37_1,SchwingerPNAS37_2,FeynmanPR84}.
Unfortunately, unlike QED and like quantum chromodynamics (QCD) at low energy, perturbation theory does not work in many-body.
Different schemes, like functional and iterative approaches, and approximations, like the GW \cite{Hedin}, have been developped along the years with some success.

The many-body QFT as we know from early works and textbooks \cite{FetterWalecka,Mahan,BaymKadanoff} is formulated in a non-relativistic form.
Relativistic effects are not negligible in condensed matter physics, as it was recognized only later.
Especially for large $Z$ atoms.
Well known manifestations are the color of gold and the liquidity of mercury \cite{rqc}.
In quantum chemistry several approximated schemes were devised to reintroduce them at the level of first quantization  \cite{rqc}.
In ordinary many-body QFT these effects are introduced in an approximate way with \textit{ad hoc} terms, like e.g. the spin-orbit.
Notice that in the ordinary theory important terms are missing both in the kinetic and in the interaction parts of the Lagrangian.
Electrons interact only by the electrostatic field instantaneously, neglecting other components of the electromagnetic field.
Moreover, fundamental symmetries like gauge invariance, crucial in particle physics QED, QCD, etc., are not checked in ordinary MBQFT.

In this letter we formulate a relativistic covariant form of many-body QFT.
We write a covariant Lagrangian density and Hamiltonian for a relativistic many-body QFT.
Eqs.~(\ref{mbl}) and (\ref{mbhd}) are the main results of this work.
To the best of our knowledge, this is the first time such a Lagrangian is presented.
The covariant many-body Lagrangian is directly derived from the QED Lagrangian integrating out internal electromagnetic field degrees of freedom by a Hubbard--Stratonovich transformation \cite{Stratonovich,Hubbard}.
MBQFT then naturally appears as a phenomenological, effective theory of the underlying QED.
This also bridges again a link between condensed matter and subnuclear physics established in early times and broken with growing specialization.
Finally we show that, under a suitable limit, the covariant Hamiltonian reduces to the ordinary many-body Hamiltonian, thus restoring the original theory.
We believe that the covariant formalism is not much more complicated than the ordinary one, both in analitical calculations and in numerical implementations into codes.
The extra-complication of Dirac 4-components, instead of Pauli 2-spinors, is largely payed back by the fact to work in an \textit{exact} formalism.
Four terms (kinetic, rest-mass, external and many-body interaction) exactly describe the physics of condensed matter systems without the need to introduce by hand further terms like the spin-orbit, the Lorentz force, the spin-spin, just to mention only few of them.

We will refer in particular to QED and to condensed matter many-body systems composed of electrons interacting electromagnetically.
But the theory can be generalized to bosons (e.g. atoms), and also to many-body systems of nucleons.
The interaction can also be of short range (e.g. nuclear).
This scheme toward an effective many-body theory can be applied to other contexts, starting for example from QCD instead of QED.
Instead of atomic, we will use \emph{natural units} where $\hbar=c=1$, $m\ne 1$, $\epsilon_0 = 1/4\pi$ and $e=\sqrt{\alpha}$.
$g_{\mu\nu}$ is the Minkowskij metric tensor and $\gamma^\mu$ are Dirac matrices.

\paragraph{Many-body covariant quantum field theory}
The formulation of any QFT starts from the expression of its Lagrangian density.
The most general Lagrangian for a many-body theory is composed of three terms,
\begin{equation}
 \lagr_\textrm{MB} = \lagr_\textrm{free} + \lagr_\textrm{ext} + \lagr_\textrm{int}
 , \label{fullmbl}
\end{equation}
where $\lagr_\textrm{free}$ is the Lagrangian for the free particles; 
$\lagr_\textrm{ext}$ accounts for the interaction of particles with an external field, taken to be classical;
finally the term $\lagr_\textrm{int}$ represents the interaction among the many particles of the system and constitutes the complication of the theory, the many-body problem to be solved.
For electrons or other fermions (represented by a Dirac field $\psi(x)$), a relativistic covariant form for $\lagr_\textrm{free}$ is the Dirac Lagrangian,
\[
 \lagr_\textrm{free}(x) = \lagr_\textrm{Dirac}(x) =
   i \bar\psi(x) \gamma^\mu \partial_\mu \psi(x) - m \bar\psi(x) \psi(x)
 ,
\]
where the first term is the kinetic and the second one is the rest mass term ($\bar\psi = \psi^\dag \gamma^0$).
The Eulero-Lagrange equation for this Lagrangian is the relativistic free-particle Dirac equation, $(i \gamma^\mu \partial_\mu - m ) \psi(x) = 0$.
For a many-particle system composed of bosons, we can take for $\lagr_\textrm{free}$ the Klein-Gordon Lagrangian
\[
 \lagr_\textrm{KG}(x) = \partial_\mu \phi^\dag(x) \partial^\mu \phi(x) - m \phi^\dag(x) \phi(x)
 ,
\]
where $\phi(x)$ can be a complex scalar field or also real in the case of neutral (e.g. atoms) bosons.

The interaction of fermions with an external electromagnetic field $A_\mu^\textrm{ext}(x)$ can be covariantly written
\begin{equation}
 \lagr_\textrm{ext}(x) = - q A_\mu^\textrm{ext}(x) \bar\psi(x) \gamma^\mu \psi(x) 
 \label{extl}
\end{equation}
($q=-e$ for electrons).
In condensed matter, whenever the Born--Oppenheimer approximation holds, the electronic degrees of freedom can be separated from the ionic ones to be integrated out.
$A_\mu^{\rm ext}(x)$ can represent the electromagnetic field generated by the ions supposed at fixed positions and seen from the many-electron system as an external field.
For example for ions on a crystal lattice $\mathbf{R}_n$, the external field will be $A^0_\textrm{ext}(x) = 1/(4\pi \epsilon_0) \sum_n Z_n e / |\mathbf{x} - \mathbf{R}_n|$ and $A^i_\textrm{ext}(x)=0$ and will be responsible of the band-structure of the solid.
We can also include into $A_\mu^{\rm ext}(x)$ a truly external electromagnetic field incident on the system.
In many-body theory only the fields associated to the (many) particles are second-quantized, while the interaction field is kept classical.
For bosons, we take the scalar electrodynamics Lagrangian
$\lagr_\textrm{ext} = -iq A_\textrm{ext}^\mu \phi^\dag \overleftrightarrow{\partial}_\mu \phi + q^2 (A_\textrm{ext})^2 \phi^\dag \phi$.
It is equally possible to write covariant forms for the interaction with other kinds of external fields.

Finally, in a relativistic theory for charged fermions interacting electromagnetically, the many-body interaction term $\lagr_\textrm{int}$ can be written in the covariant form
\begin{equation}
 \lagr_\textrm{int}^\textrm{MB}(x) = - \frac{1}{2} \int dy \, 
    j^\mu(x) D_{\mu\nu}(x,y) j^\nu(y) 
 \label{mbl}
\end{equation}
where $j^\mu(x) = q \bar\psi(x) \gamma^\mu \psi(x)$ is the electron current and
\begin{eqnarray}
 && D^{\mu\nu}(x,y) = \int \frac{dk}{(2\pi)^4} e^{-ik(x-y)} \, \tilde{D}^{\mu\nu}(k)
 , \nonumber \\
 && \tilde{D}^{\mu\nu}(k) = \frac{1}{\epsilon_0} \frac{- g^{\mu\nu}}{k^2 + i\eta}
 ,\label{phprop}
\end{eqnarray}
is the photon propagator in real and reciprocal space (under the Lorentz condition). 
This expresses the interaction of two quadri-currents $j^\mu$ associated to fermions with charge $q$.
We can write a similar term for charged bosons interacting electromagnetically.
Other possible interactions may be considered, for example short-range (nuclear) interactions or even a local interaction of the form $\lagr_\textrm{int}(x) = - \lambda \phi^4(x)$.

Finally, the covariant many-body Hamiltonian density is $\hamil = \pi \dot{\psi} + \bar\pi \dot{\bar\psi} - \lagr$, where $\pi = \partial \lagr / \partial \dot{\psi}$ and $\bar{\pi} = \partial \lagr / \partial \dot{\bar{\psi}}$ are the canonically conjugated fields,
\begin{eqnarray}
 && \hamil_\textrm{MB}(x) = \bar\psi(x) (-i \gamma^i \partial_i + m) \psi(x) + j^\mu(x) A^\textrm{ext}_\mu(x) + 
  \nonumber \\
  && + \frac{1}{2} q^2 \int dy \, \bar{\psi}(x) \gamma^\mu \psi(x) D_{\mu\nu}(x,y) \bar{\psi}(y) \gamma^\nu \psi(y)
 . \label{mbhd}
\end{eqnarray}
The Hamiltonian is obtained by integrating $\hamil_\textrm{MB}(x)$ over the space,
\begin{equation}
  H_\textrm{MB}(t) = \int d\mathbf{x} \, \hamil_\textrm{MB}(x)
  .  \label{hmb}
\end{equation}

\paragraph{Derivation of the many-body Lagrangian from QED}
We will now show how the many-body Lagrangian $\lagr_\textrm{MB}$ can be derived from the QED Lagrangian.
Let us suppose for the moment that our many-body system is composed only by electrons, without ions.
We start from the QED Lagrangian for electrons,
\[
 \lagr_\textrm{QED} = \lagr_\textrm{Dirac} + \lagr_\textrm{Maxwell} 
\]
where the Maxwell Lagrangian is $\lagr_\textrm{Maxwell} = -1/4 F^{\mu\nu} F_{\mu\nu} - j_\mu A^\mu$, in terms of the electromagnetic field $A^\mu$ and tensor $F^{\mu\nu} = \partial^\mu A^\nu - \partial^\nu A^\mu$. This is equivalent, up to a total divergence of a quadrivector, to
\[
 \lagr_\textrm{Maxwell} = \frac{1}{2} A^\mu ( g_{\mu\nu} \partial^\lambda \partial_\lambda - \partial_\mu \partial_\nu ) A^\nu  - j_\mu A^\mu
 ,
\]
composed of a first free electromagnetic field term and a second term describing the interaction of the electromagnetic field $A^\mu$ with a charged current $j_\mu$, via a minimal substitution $i \partial_\mu \to i \partial_\mu - q A_\mu$ (covariant derivative) into $\lagr_\textrm{Dirac}$.
The Eulero-Lagrange equations associated to $\lagr_\textrm{Maxwell}$ are the Maxwell equations, $\partial^\lambda \partial_\lambda A^\mu(x) - \partial^\mu \partial_\nu  A^\nu(x) = j^\mu(x)$.
For the system without ions, the sources $j^\mu$ of the electromagnetic field are only associated to the many-electrons of the system.
We now write the action $S$ and the generating functional $Z$ associated to the Maxwell Lagrangian,
\[
 Z[A,j] = e^{i S[A,j]} = e^{i \int dx \lagr_\textrm{Maxwell}[A,j](x)}
 .
\]
In many-body the degrees of freedom associated to the interaction fields are integrated out toward an effective generating functional/Lagrangian where only matter fields appear,
\[
 Z_\textrm{eff}[j] = \int \mathscr{D}A \, \, Z[A,j] = e^{i S_\textrm{eff}[j]} = e^{i \int dx \lagr_\textrm{eff}[j](x)}
 .
\]
The functional integral $\int \mathscr{D}A$ over the electromagnetic field  is performed by a Stratonovich transformation \cite{Stratonovich}
\[
 \int \mathscr{D}\phi \, \, e^{- \frac{1}{2} \phi O \phi + s \phi} = \textrm{const} \cdot e^{\frac{1}{2} s O^{-1} s}
\]
By replacing $\phi$ with $A^\mu$, $O$ with $- ( g_{\mu\nu} \partial^\lambda \partial_\lambda - \partial_\mu \partial_\nu )$ and $s$ with $-j_\mu$, we get
\[
 Z_\textrm{eff}[j] = e^{i \int dx dy (-) \frac{1}{2} j^\mu D_{\mu\nu} j^\nu}
 ,
\]
where $D_{\mu\nu} = ( g_{\mu\nu} \partial^\lambda \partial_\lambda - \partial_\mu \partial_\nu )^{-1}$ is the inverse (Green's) function of the Maxwell operator (the photon propagator, if $A^\mu$ were quantized).
The effective Lagrangian reads
\[
 \lagr_\textrm{eff}[j](x) = - \frac{1}{2} \int dy \, j^\mu(x) D_{\mu\nu}(x,y) j^\nu(y)
 ,
\]
which corresponds exactly to the covariant many-body interaction Lagrangian, Eq.~(\ref{mbl}).
The 2-body interaction (pairing two $j^\mu$) presented by Eq.~(\ref{mbl}) is due to the Abelian character of QED.
A non-Abelian theory would lead also to further-order-body interaction terms.

If we want to introduce the effect of the ions or a more general external electromagnetic field associated to external sources, thank to the superposition principle we can split $A^\mu = A_\textrm{ext}^\mu + A_\textrm{int}^\mu$ into an external part $A^\mu_\textrm{ext}$, whose sources are the external currents and ions, and an internal $A_\textrm{int}^\mu$, due to the electrons.
Only $A_\textrm{int}^\mu$ is integrated out by a Stratonovich transformation, while $A^\mu_\textrm{ext}$ is a degree of freedom of the final effective Lagrangian.
Beyond the interaction term Eq.~(\ref{mbl}), we will also get a term $\lagr_\textrm{ext} = - j_\mu A_\textrm{ext}^\mu$, corresponding to the external Lagrangian Eq.~(\ref{extl}), plus another term associated to the equation of motion of $A^\mu_\textrm{ext}$ that we can drop since we regard this field as non-dynamical and fully determined by the ion positions and motions and eventual external sources.
This concludes our derivation of the full many-body Lagrangian $\lagr_\textrm{MB}$, Eq.~(\ref{fullmbl}).

\paragraph{Relation QED -- Many-Body QFT}
The previous derivation shows that QED is the theory underlying many-body QFT.
However the latter is not simply a non-relativistic limit of QED.
A covariant relativistic many-body theory, with respect to QED, should be rather regarded as an effective, phenomenological theory, where the degrees of freedom associated to the internal (non quantized) electromagnetic field are integrated out.
The many particles directly interact by the factor $D_{\mu\nu}$ accounting for the effect of the intermediate field, without introducing it explicitly.

QED deals with scattering of electrons coming into interaction for short times --- thus exchanging only few photons before leaving again to far distances --- and where also photons appear as initial or final scatterers.
So in QED the electromagnetic field plays a central role, and it is second-quantized like matter fields.
Many-body QFT deals with a different problem: the organization of many electrons, exchanging lots of photons, in order to form bound states and condensed matter systems.
Here the electromagnetic field keeps classical; not even the need to introduce it explicitly:
the many-body interaction between electrons can be described as direct.
Of course, one can try to tackle the many-body problem directly from QED, following the dynamics of every single electron interacting by the exchange of a large number of photons.
The difficulty is evident.
In that respect, the relation between QED and MBQFT is akin to the one between classical and statistical mechanics.


We stress that MBQFT starts from already QED-renormalized coupling constant and masses.
Both theories at the end face the same difficulty, namely the treatment of the interaction term, $j^\mu A_\mu$ in QED and $j^\mu D_{\mu\nu} j^\nu$ in MBQFT.

\paragraph{Reduction to the ordinary many-body Hamiltonian}
We will now show that the covariant many-body Hamiltonian reduces, under more restrictive hyphotheses and in the non-relativistic limit, to the ordinary many-body Hamiltonian used in condensed matter \cite{FetterWalecka,Mahan,BaymKadanoff}.

In the non-relativistic limit, the Eulero-Lagrange equations of $\lagr_\textrm{Dirac} + \lagr_\textrm{ext}$ give the Pauli equation for an electron into an electromagnetic field $A_\textrm{ext}^\mu(x)$ \cite{ItzyksonZuber}.
The 4-components Dirac spinor is rewritten as $\psi(x) = e^{-imx^0} \big( \phi(x) \, \chi(x) \big)$, singling out an oscillatory factor $e^{-imt}$ correspondent to the largest energy in play in the non-relativistic limit (the rest mass $m$), and splitting $\psi$ into two 2-components spinors: the upper $\phi$ and the lower $\chi$, slowly varying functions of $t$.
$\phi$ and $\chi$ correspond to, respectively, positive and negative energy solutions of the Dirac equation for an electron.
In the non-relativistic limit of kinetic energies small in comparison to $m$, $\chi$ reveals $v/c$ smaller than $\phi$, and can be neglected.
This leaves $\phi$ obeying the Pauli equation for an electron into an electromagnetic field $A_\textrm{ext}^\mu(x)$.
Our Hamiltonian contains not only the interaction of the electron with an electric field, but also the interaction of its orbital and spin magnetic moment with a magnetic field.
A more systematic way to pursue this limit is the Foldy-Wouthuysen transformation \cite{FoldyWouthuysen} providing other three important terms: the relativistic velocity correction $p^4/8m^3$, the Darwin, and the spin-orbit terms.

Let's now focus on the many-body interaction last term of the Hamiltonian.
The photon propagator Eq.~(\ref{phprop}) can be split in three components \cite{MandlShaw},
$D = D_\textrm{T} + D_\textrm{C} + D_\textrm{R}$,
where $\tilde D_\textrm{T}^{\mu\nu}(k)=\epsilon_0^{-1} \sum_{r=1}^2 \varepsilon_r^\mu \varepsilon_r^{\nu}/k^2$ is the transverse photon propagator with the polarization vectors $\varepsilon_r^\mu(k)$ chosen as $\boldsymbol{\varepsilon}_r \perp \mathbf{k}$ (or $\varepsilon_r^\mu k_\mu = 0$) and in two $r=1,2$ orthogonal space directions. $\tilde D_\textrm{R}^{\mu\nu}(k)$ is the component proportional to either $k^\mu$ or $k^\nu$ or both. This component gives no contribution when coupled to conserved currents $j^\mu$, $\partial_\mu j^\mu(x) = 0$, so that it does not contribute to the many-body Lagrangian or Hamiltonian Eqs.~(\ref{mbl}, \ref{mbhd}).
Finally the component along time directions,
\[
 \tilde D_\textrm{C}^{\mu\nu}(k) = \frac{1}{\epsilon_0} \frac{g^{\mu 0} g^{\nu 0}}{\mathbf{k}^2}
\]
is defined as the Coulomb propagator, as we will see.
If now in Eqs.~(\ref{mbhd}, \ref{hmb}) we neglect the transvers propagator $D_\textrm{T}$ and assume the approximation $D \simeq D_\textrm{C}$
\[
  D_\textrm{C}^{\mu\nu}(x) = \int \frac{dk}{(2\pi)^4} e^{-ikx} \frac{1}{\epsilon_0} \frac{g^{\mu 0} g^{\nu 0}}{\mathbf{k}^2} = 
   \frac{g^{\mu 0} g^{\nu 0} \delta(x^0)}{4\pi \epsilon_0 |\mathbf{x}|}
 ,
\]
we are left with the Hamiltonian
\footnote{The same result can also be obtained fixing the Coulomb gauge $\boldsymbol{\nabla} \cdot \mathbf{A} = 0$ on the photon propagator \cite{Landau} and neglecting its spatial $D^{ij}$ components.}
\[
 H_{\rm int}(t) = \frac{1}{2}  \frac{q^2}{4\pi \epsilon_0} 
   \int d\mathbf{x} d\mathbf{y} 
     \frac{\psi^\dagger(\mathbf{x},t) \psi(\mathbf{x},t) \psi^\dagger(\mathbf{y},t) \psi(\mathbf{y},t)}
          {|\mathbf{x}-\mathbf{y}|}
 ,
\]
which can be recognized as the ordinary many-body interaction Hamiltonian, starting point of many-body textbooks \cite{FetterWalecka,Mahan,BaymKadanoff}.
In this Hamiltonian electrons interact only via the instantaneous electrostatic Coulomb interaction.
\textit{This is an approximation with respect to the covariant Hamiltonian} $H_\textrm{MB}$, Eq.~(\ref{hmb}), which contains $D_\textrm{T}$ too.

\paragraph{Advantages of the covariant theory}
The covariant Hamiltonian Eq.~(\ref{mbhd}, \ref{hmb}) contains not only the relativistic correction to the velocity, the spin-orbit and the Darwin terms; but also many-body \textit{magnetic}, or better \textit{electromagnetic} (beyond electrostatic-only) interactions between the electrons.
These comprise for example the Lorentz force exerced on a given electron and due to the magnetic field generated by other electrons in motion with respect to the first.
As well as many-body interactions where also the spin of the electron plays a role, e.g. many-body spin-spin or spin of the first electron - orbit of the second, etc.
Moreover, the exact many-body interaction is non-instantaneous and contains retardation effects.
In the ordinary theory, many-body interactions beyond the pure instantaneous electrostatic interaction are usually neglected, or introduced by hand whenever the approximation reveals too crude.
The spin also is introduced by hand, together with the long list of terms where it appears, e.g. the spin-orbit coupling term, the coupling with an external magnetic field or in spin-spin, spin-orbit many-body interaction terms.
In the relativistic theory the spin is naturally introduced.
\textit{The Hamiltonian Eq.~(\ref{mbhd}) contains in a compact way all the correct terms and it is exact.}

\paragraph{Gauge invariance and causality}
The covariant Lagrangian Eq.~(\ref{mbl}) owns another important characteristics: it is \textit{gauge invariant}.
It can be verified that under the gauge transformation:
\begin{eqnarray*}
 && A_\mu^\textrm{ext}(x) \to A_\mu^{' \textrm{ext}}(x) = A_\mu^\textrm{ext}(x) + \partial_\mu \lambda(x) \\
 && \psi(x) \to \psi'(x) = e^{i q \lambda(x)} \psi(x)
\end{eqnarray*}
the sum $\lagr_\textrm{Dirac} + \lagr_\textrm{ext}$ is unchanged.
Under a gauge transformation, the photon propagator transforms
$\tilde D_{\mu\nu} \to \tilde D'_{\mu\nu} = \tilde D_{\mu\nu} + \lambda_\mu k_\nu + \lambda_\nu k_\nu$
\cite{Landau} by terms proportional to $k_\mu$ that give no contributions when coupled to conserved currents $j^\mu$: $\lagr^\textrm{MB}_\textrm{int}$ is invariant.

Together with Lorentz invariance and unitarity, causality is another important requirement met by the many-body S-matrix approach discussed in this letter.
Causality conditions can constraint physical amplitudes in quantum theory \cite{causality}.
They are strictly related to the vanishing bosonic (fermionic) field commutator (anticommutator) outside the light-cone, which simply restates that two events separated by a space-like interval cannot interfere.
After the Stratonovich transformation, the fermionic field operators are second-quantized and their anticommutativity outside the light-cone is naturally imposed.
By this choice, dispersion relations can be assessed for the relevant phenomenological quantities (physical conditions for renormalization, analytical properties of polarization/self-energy operators, Kallen--Lehmann representation for propagators, and so on).

\paragraph{Extensions}
The idea underlying our scheme has the potentiality of generalization to other contexts.
Any theory where the strenght/vector field looks classical because of the large number of quanta involved in the relevant phenomenology among the second-quantized matter fields, can be projected onto a many-body effective theory by integrating out the vector field.
For instance, non-Abelian field theories (effective quantum chromodynamics for color-transparent bound states and, in turn, nuclear physics), will benefit of such an effective scheme, once provided a viable Stratonovich transformation for the triple/quartic couplings, typical of such theories.

\paragraph{Applications}
The present relativistic covariant formalism is not much more complicated than the ordinary many-body formalism.
In the many-body term of $\hamil_\textrm{MB}$ Eq.~(\ref{mbhd}) we need only to replace the Coulomb interaction $v_\textrm{C}=1/|\mathbf{x}-\mathbf{y}|$ by a photon propagator $D_{\mu\nu}(x,y)$.
The external term is generalized to include also interaction with the magnetic field.
The gain is that, with the same terms, we also take into account effects like the spin-orbit interaction or the Lorentz force, that should be normally introduced by hand with \textit{ad hoc} further terms in ordinary many-body.
The only extra-complication is to deal with Dirac 4- instead of Pauli 2-components spinors.
In a first quantization formalism the added negative energy solutions associated to anti-matter can constitute a problem 
\cite{Landau}.
But this is not the case in a second-quantized quantum field theory, like MBQFT.
Many-body analytical calculations and numerical codes can be easily generalized to a covariant formalism, without major problems.

\paragraph{Conclusions}
We have presented a relativistic covariant form of many-body theory derived from QED and shown to reduce to ordinary many-body theory.
The resulting formalism is compact and more complete than the ordinary.
It is moreover causal and gauge-invariant.



\begin{thebibliography}{99}

\bibitem{FetterWalecka}
A. L. Fetter and J. D. Walecka, \textit{Quantum Theory of Many-Particle Systems}, McGraw-Hill 1971. (Eq.~(2.4)).

\bibitem{Mahan}
G. D. Mahan, \textit{Many-Particle Physics}, Plenum 1980.

\bibitem{BaymKadanoff}
L.~P.~ Kadanoff and G.~Baym, \textit{Quantum Statistical Mechanics}, Benjamin, New York 1962. (Eq.~(1-2)).

\bibitem{Hedin}
L.~Hedin, Phys. Rev. {\bf 139}, 796 (1965).

\bibitem{DysonPR75_1}
F.~J.~Dyson, Phys. Rev. {\bf 75}, 486 (1949).

\bibitem{DysonPR75_2}
F.~J.~Dyson, Phys. Rev. {\bf 75}, 1736 (1949).

\bibitem{SchwingerPNAS37_1}
J.~Schwinger, Proc. N.A.S. {\bf 37}, 452 (1951).

\bibitem{SchwingerPNAS37_2}
J.~Schwinger, Proc. N.A.S. {\bf 37}, 455 (1951).

\bibitem{FeynmanPR84}
R.~P.~Feynman, Phys. Rev. {\bf 84}, 108 (1951).

\bibitem{rqc}
M. Reiher and A. Wolf, \textit{Relativistic Quantum Chemistry}, Wiley, Weinheim 2009.

\bibitem{Stratonovich}
R. L. Stratonovich, Soviet Phys. Dokadly \textbf{2}, 416 (1958).

\bibitem{Hubbard}
J. Hubbard, Phys. Rev. Lett. \textbf{3}, 77 (1959).

\bibitem{ItzyksonZuber}
C. Itzykson and J.-B. Zuber, \textit{Quantum Field Theory}, McGraw-Hill 1980. In particular Sec.s 2-2-3 and 2-2-4.

\bibitem{MandlShaw}
F. Mandl and G. Shaw, \textit{Quantum Field Theory}, Wiley 1990. In particular Sec. 5.3.

\bibitem{Landau}
V.~B.~Berestetskii, E.~M.~Lifshitz and L.~P.~Pitaevskii, \textit{Theoretical Physics}, Vol. IV, Pergamon 1982. Sec.~76.


\bibitem{FoldyWouthuysen}
L. L. Foldy and S. A. Wouthuysen, Phys. Rev. \textbf{78}, 29 (1950).

\bibitem{causality}
M.~Gell-Mann, M.~L.~Goldberger and W.~E.~Thirring, Phys. Rev. \textbf{95}, 1612 (1954).

\end{thebibliography}
\end{document}